\documentstyle[pre,twocolumn,aps,epsf]{revtex} 

\begin{document}

\title{Dynamical properties of the Zhang model of Self-Organized
Criticality}

\author{Achille Giacometti\cite{achillemail}}
\address{INFM Unit\'a di Venezia and
Dipartimento di Scienze Ambientali,
Universita' degli Studi di Venezia,
I-30123 Venezia, Italy}
\author{Albert D\'{\i}az-Guilera\cite{albertmail}}
\address{
Departament de F\'{\i}sica Fonamental, 
Universitat de Barcelona, 
Diagonal 647, E-08028 Barcelona, Spain}

\date{\today}

\maketitle

\begin{abstract}
Critical exponents of the infinitely slowly
driven Zhang model of self-organized criticality 
are computed
for $d=2,3$ with particular emphasis devoted
to the various roughening exponents.
Besides confirming recent estimates of
some exponents, new quantities are monitored
and their critical exponents computed.
Among other results, it is shown
that the three dimensional exponents
do not coincide with the Bak, Tang,
and Wiesenfeld (abelian) model and that
the dynamical exponent as computed from
the correlation length and
from the roughness of the energy profile do not
necessarily coincide as it is
usually implicitly assumed. An explanation
for this is provided. The possibility
of comparing these results with
those obtained from Renormalization Group
arguments is also briefly addressed.
\end{abstract}


\section{Introduction}
\label{sec:introduction}
Despite more of a decade of intensive studies,
the phenomenon named Self-Organized Critically (SOC)
by Bak, Tang, and Wiesenfeld (BTW) \cite{prl59.381}
is far from being fully understood.
The name SOC originates from the fundamental
property that
an open system externally driven in
a (infinitely) slow fashion, settles into
a critical state with no characteristic time
and length scales, without any parameter tuning;
see e.g. Ref. \cite{Grinstein96} for a review.

Although a large number of
recipes have been proposed as toy models
to mimic this behavior, the original sandpile
model \cite{prl59.381} still carries most of the
information presented in this phenomenon.
A variation of this model was introduced
a couple of years later by Zhang \cite{prl63.470}.
The basic differences with respect to the BTW model were:
first, the variable describing the state of the lattice site 
could take
continuous rather than discrete values;
second, the BTW model
is {\em abelian} \cite{prl64.1613} while the Zhang model is not.
In spite of this differences, extensive recent numerical
simulations \cite{Lubeck97} on the two-dimensional Zhang model,
opened the possibility that they both belong to
the same universality class, in disagreement with
the original scaling prediction by Zhang \cite{prl63.470}. 
Apart from the aforementioned investigation \cite{Lubeck97},
the Zhang model was already studied in different dimensionalities
in Ref. \cite{pra42.769} where an estimates for
some critical exponents, notably the avalanche size exponent 
$\tau_s$, was given. 
However, these estimates, whose main aim
was to test the robustness of universality
of the model under anisotropy of the energy repartition,
appeared to be based on small sizes and statistics. 

On the other hand, a Langevin counterpart of the
Zhang model was repeatedly studied by Renormalization 
Group (RG) methods \cite{pra45.8551,f1.963,Ghaffari96,Corral97}, and
predictions for critical exponents
in a one-loop working scheme were drawn. 
The dynamical exponent $z$ as calculated from the
correlation function in the case when the additive noise has a 
typical time
scale much bigger than the relaxation time scale \cite{f1.963}, 
turned out to be very close to the one relating the
correlation length and the relaxation time in the standard
dynamical scaling hypothesis \cite{macp} in the Zhang model.

It is then desirable to have a more complete numerical
investigation touching upon those issues appearing
in the RG calculations and those which were previously neglected.
This is indeed the aim of the present work where a fairly
complete analysis of the model, in different dimensionalities is
carried out and compared, when possible, with previous numerical
and RG work. By doing this we found few unexpected results.

Firstly, the three-dimensional results do not support the conjecture
that the Zhang and the BTW models belong to the same universality 
class.
Secondly, whereas it is true that the exponent $z$ of the
Zhang model is very close to the one obtained by RG techniques
as previously discussed, the roughening exponent is not \cite{f1.963}.
Finally, the critical exponent $z$ is {\em different}
when calculated from the dynamical scaling {\em ansatz}
and when computed from the roughness exponent.
This latter discrepancy can be fixed in our case by noting
that the correlation length (maximum avalanche distance)
does {\it not} scale linearly with system size $L$.

The plan of the paper is as follows. In Sec.~\ref{sec:model}
the model is defined, whereas in Sec.~\ref{sec:quantities}
all relevant quantities concurring to identify the critical
behavior of the model  are laid down. Sec.~\ref{sec:results}
contains the results of this effort and comparisons with earlier
ones,
and finally some conclusive remarks are left in 
Sec.~\ref{sec:conclusions}.

\section{The slowly driven Zhang model}
\label{sec:model}
Each point of an hypercubic lattice is characterized
by a {\em continuous} energy variable $E_{\tau}({\bf x},t)$, where 
${\bf x}$
denotes the lattice position, $t$ the driving (slow) time, and
$\tau$ the relaxation (fast) time. Whereas $t$ runs from 0 to a
sufficiently large value needed to get good statistics, $\tau$ 
runs
from 0 to $T(t)$, which is the total fast time that an
avalanche initiated at a slow time $t$ takes to be completed.
In this way the two time scales are well separated. Starting from an
initially empty lattice, the
dynamics of the evolution is defined as follows \cite{prl63.470}:
\begin{enumerate}
\item Start with a randomly chosen lattice point ${\bf x}_0$
and set it slightly above some critical energy $E_c$ (hereafter
chosen to be $1$ without loss of generality) by repeated addition
of a random energy taken uniformly 
from the interval $(0,1/4)$ \cite{note1,cafiero}. 
\item The site ${\bf x}_0$ relaxes according to
the following equation:
\begin{eqnarray} \label{relaxation_slow}
E_{\tau+1}({\bf x},t) &=&
[1-\theta(E_{\tau}({\bf x},t)-E_c)]E_{\tau}({\bf x},t) + \\
&+& \frac{1}{2d}
\sum_{{\bf y}({\bf x})} 
\theta(E_{\tau}({\bf y},t)-E_c)\;E_{\tau}({\bf y},t)
\nonumber
\end{eqnarray}
where $\theta(\cdot)$ is the Heaviside step function and $d$
is the space dimension. Here 
the notation $\sum_{{\bf y}({\bf x})}$ means that the
sum is restricted to the nearest-neighbors ${\bf y}$ of
site ${\bf x}$. Clearly this is tantamout to say that
each site ${\bf x}$ whose energy exceeds a critical value
$E_c$ is set to zero and its energy is equally redistributed
to the nearest-neighbors. 
\item Iterate Step 2 for the other sites that become critical until
all sites are below $E_c$.
\item At this point increase $t$ by one unit ($t \rightarrow t+1$),
and randomly pick
a new initial seed ${\bf x}_0^{\prime}$ in Step 1.
\end{enumerate}
The process is iterated until the system has reached
a {\em steady-state} configuration where the average energy
\begin{eqnarray} \label{E_ave}
\overline{E(t)} &=& \frac{1}{V} \sum_{{\bf x}} E({\bf x},t)
\end{eqnarray}
reaches a well defined value.
Here $V=L^d$ is the volume of the lattice. We note that whenever
there is no
subscript for the energy it will be implicitly assumed that the
avalanche is over, i.e. $\tau$ has reached $T(t)$. Starting at this
time, when the system has reached a stationary state, we collect 
all the relevant dynamical properties.

\section{Probability distributions and correlation functions}
\label{sec:quantities}
At each time $t$ there is a growing avalanche; within the fast time
scale we can measure the number of active sites at each update
($\tau$)
\begin{equation}
S_{\tau}(t) = \sum_{{\bf x}} \theta(E_{\tau}({\bf x},t)-E_c)
\end{equation}
and from this we can define the size of an avalanche at time $t$
\begin{equation}
S(t)=\sum_{\tau=1}^{T(t)}S_{\tau}(t).
\end{equation}
From the size of the avalanche we can compute a characteristic length
$\xi(t)$ defined as the radius of gyration with respect to the seed 
site
${\bf x}_0$.
This characteristic
length is related to the time the avalanche needs to be completed
through the standard relation \cite{macp}
\begin{eqnarray} \label{def_z}
T(t) \sim \xi^z(t)
\end{eqnarray}
which defines the dynamical exponent $z$.

Other quantities that are interesting to measure are the total input
and output currents. They are defined as follows
\begin{eqnarray} \label{input_current}
J_{\mbox{\scriptsize{in}}}(t) &=& \delta E({\bf x}_0,t)
\end{eqnarray}
\begin{eqnarray} \label{output_current}
J_{\mbox{\scriptsize{out}}}(t) &=& \sum_{\tau=0}^{T(t)} 
\sum_{{\bf x} \in \partial
\Lambda} E_{\tau}({\bf x},t) 
\end{eqnarray}
where $\Lambda$ is the bulk and $\partial \Lambda$ is the boundary of
the bulk (the sum of the two forming the total available lattice 
space).
Here $\delta E({\bf x}_0,t)$ is the total added energy necessary to
the site to be active (i.e. above the critical energy $E_c=1$).

In order to take into account the existence of two different time
scales one should be very careful when defining the correlation
functions.
Upon extending (\ref{E_ave}), we can define 
the $q-$th spatial moment of the energy as
\begin{eqnarray} \label{q_moment}
\overline{E^q}(t) &=& \frac{1}{V} \sum_{{\bf x}}
E^q({\bf x},t)
\end{eqnarray}
and then the interface width (or roughness) \cite{barabasi}
is
\begin{eqnarray} \label{roughness_slow}
W_s(t,L) &=&
\sqrt{\overline{E^2(t)}-\overline{E(t)}^2.
}
\end{eqnarray}
This definition applies to the {\it slow} time
scale as also indicated by the suffix $s$,
and coincides with the usual definition
of roughness in the framework of growth processes.
On the other hand one could think to measure the
energy fluctuations {\it during} the evolution
of an avalanche. Since an avalanche of duration
$\tau$ occurs at many different input times $t$,
we define the following {\it fast} roughness:
\[
W_f^2(\tau,L)=
\]
\begin{equation}
=< \frac{1}{V} \sum_{{\bf x}}
E_{\tau}^2({\bf x},t)-\left(\frac{1}{V}
\sum_{{\bf x}}E_{\tau}({\bf x},t)\right)^2 >_{t}.
\label{roughness_fast}
\end{equation}
In Eq. (\ref{roughness_fast}) the roughness is averaged
over
different times $t$ (and hence avalanches).

According to standard scaling hypothesis 
(see e.g. \cite{barabasi}) one expects these correlation
functions to display the following scaling forms:
\begin{mathletters}
\begin{equation} \label{scal_w_fast}
W_f(\tau,L)= \tau^{-\beta_f} \Phi_f(\tau /L^{z_f})
\end {equation}
\begin{equation} \label{scal_w_slow}
W_s(t,L)= t^{\beta_s} \Phi_s(t /L^{z_s})
\end {equation}
\label{eqs:scaling}
\end{mathletters}%
where $\Phi_f$ anf $\Phi_s$ are finite size functions.

In (\ref{scal_w_fast}) 
the roughness is expected to {\it decrease} 
rather than to {\it increase} as in more
conventional growth processes \cite{barabasi},
because the maximum energy is bounded
and the avalanche is a relaxational process. 

\section{Critical exponents and Results}
\label{sec:results}

This model was already carefully investigated in two-dimensions. 
Apart from the original work \cite{prl63.470}, recent extensive simulations
on remarkably large sizes were carried out in $d=2$\cite{Lubeck97}.
When comparable, our results are in good agreement with
both previous analysis.
However, in these papers, the behavior of some important quantities,
necessary to our purposes, were neglected,
nor was a complete study in dimensionality different from $2$ ever attempted.
Indeed while Zhang only reported the steady state value of the
average energy along with the "quantized" energy distribution
$P(E)$ for $d=3$, in \cite{pra42.769} a value for the avalanche
size exponent (see below) is reported for dimensions up to $4$.
The latter was however probably 
based on very small sizes without any attempt of a finite scale
analysis.
As a result these estimates, albeit close, turn out to be slightly 
off
compared to ours.

In our simulations we used sizes up to $L=300,60$ 
and times up to $2^{17},2^{18}$
in $d=2,3$ respectively. These are smaller than the
ones used in \cite{Lubeck97} for $d=2$ but considerably larger than 
all other three dimensional studies.

For the sake of clarity and compactness, let us now review
some known results first.
As it is well known by now, the system reaches a steady state (where
the average energy is no longer changing) after a transient which
clearly scales as $L^d$ since it takes that many time steps (on
average) to "explore" the whole lattice. 
The resulting values of the stored energy
$\overline{E(t)}$ are $0.63 \pm 0.01$ and $0.58 \pm 0.01$
(estimated from the largest sizes) in $d=2,3$ respectively.
These results are in agreement with those found by
Zhang in his original simulations.
Another feature already observed by Zhang is that the critical state
has energy which is peaked around well defined energies the number of
which depends only on the dimensionality of the hypercubic lattice.
It has also been established that this feature is unchanged upon
introducing
an asymmetry in the probability distribution and by introducing
different lattices \cite{pra42.769}.

\begin{figure}
\epsfxsize=3.8truein\epsffile{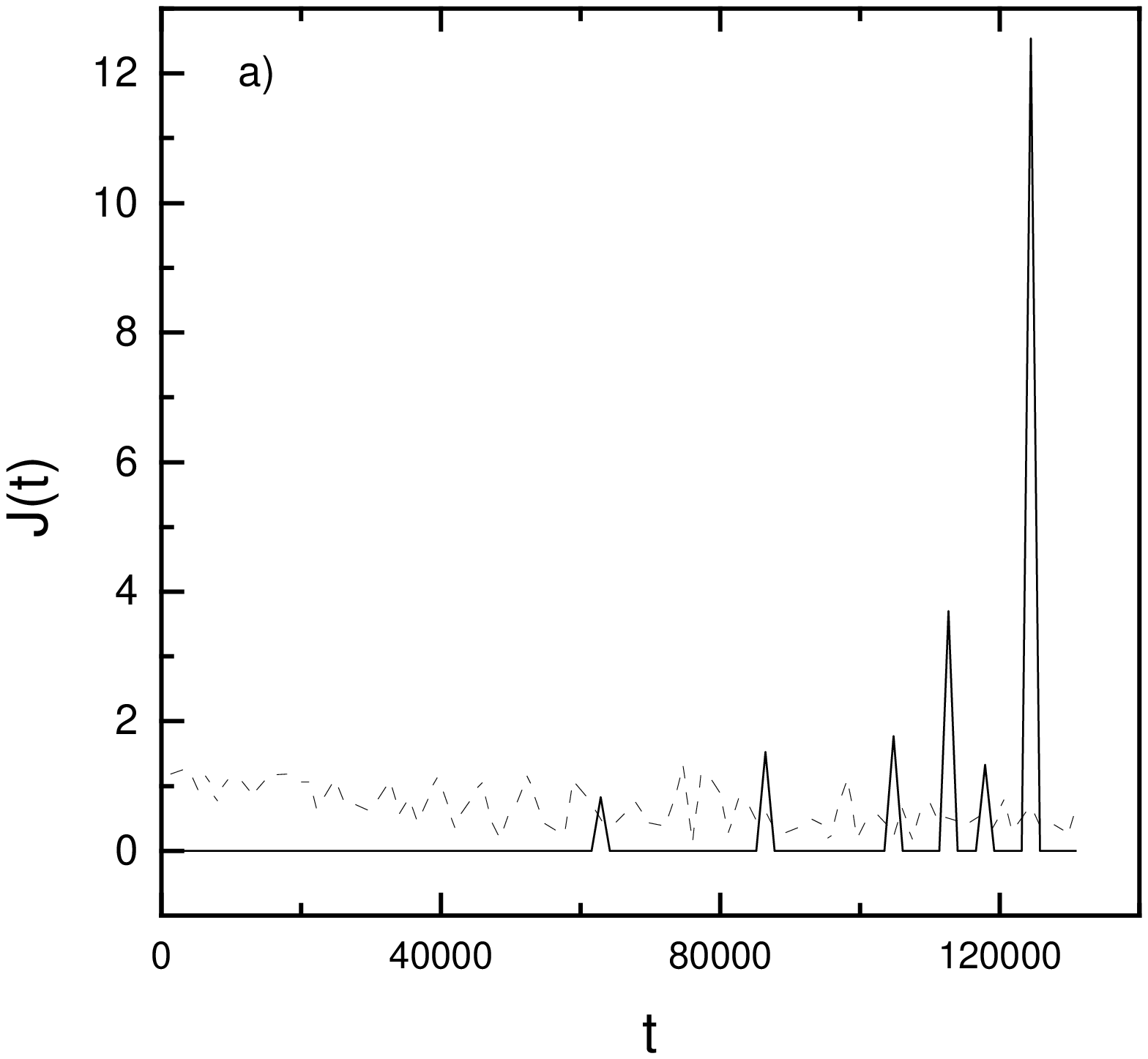}
\epsfxsize=3.8truein\epsffile{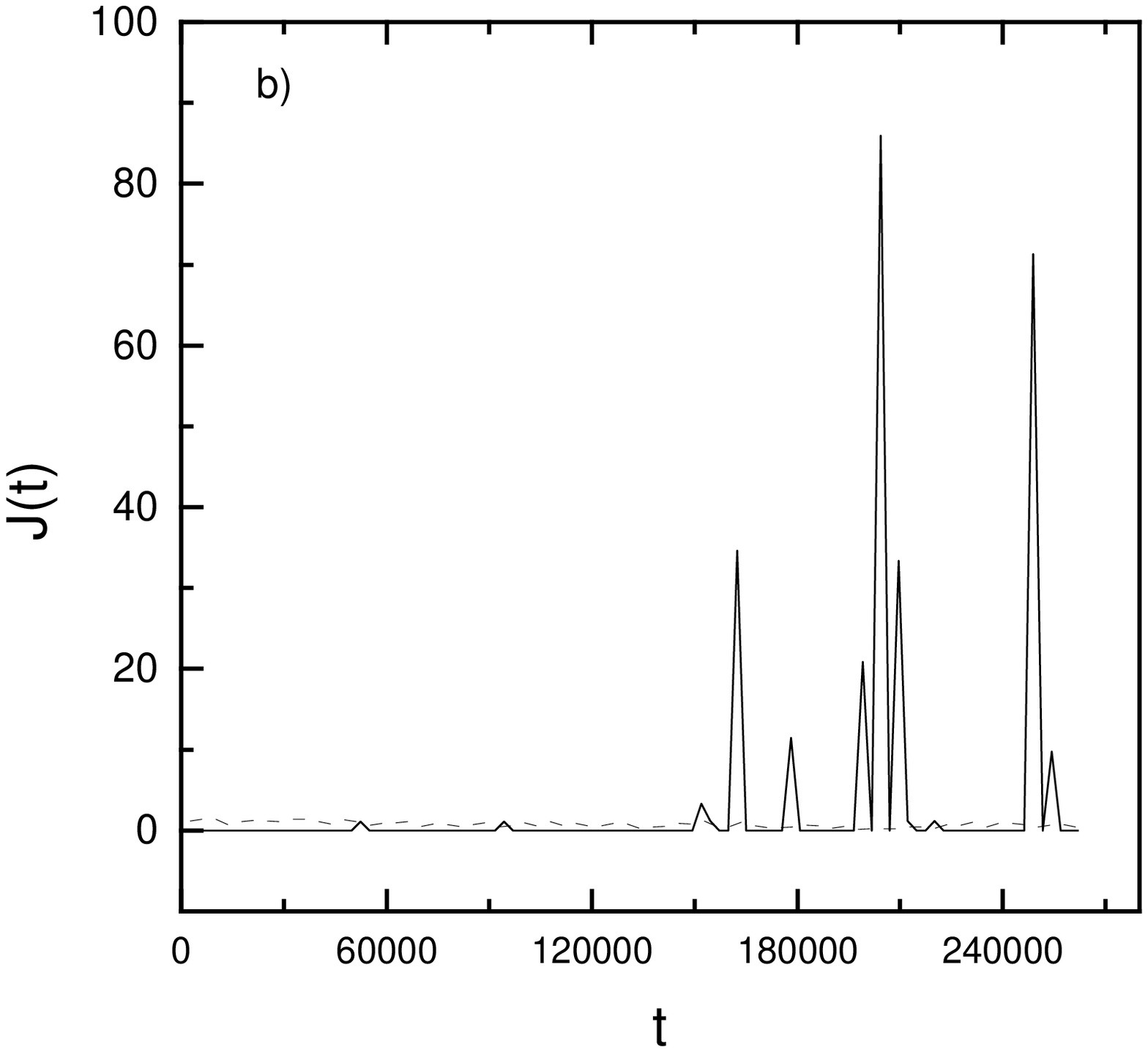}
\caption{Plot of the total energy $J(t)$, both in (dotted line) 
and out (full line), in $d=2$ (a) and $d=3$ (b).}
\label{fig1}
\end{figure}

As explained by Hwa and Kardar \cite{pra45.7002} in the framework 
of the
one-dimensional BTW sandpile model, monitoring the total
output energy current proves to be very useful in
understanding the
mechanism that leads to the steady state. 
This is shown in Fig.~\ref{fig1}.
Whereas clearly the input current
is a random function between 0 and 1, the output current displays
sequences
of bursts followed by long periods of quiescence similar to
the one found by Hwa and Kardar in the slow driving regime. 
We also computed its power spectrum
$S({\nu})$ (the Fourier transform of the output current-current
correlation) which appears to be white noise in all cases. This 
is related
with the fact that our system corresponds to a {\em non-interacting
avalanche regime} in their language \cite{pra45.7002}.

\begin{figure}
\epsfxsize=3.8truein\epsffile{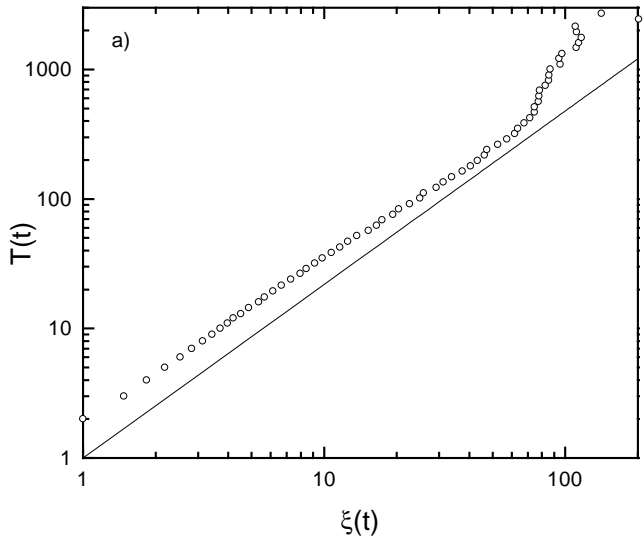}
\epsfxsize=3.8truein\epsffile{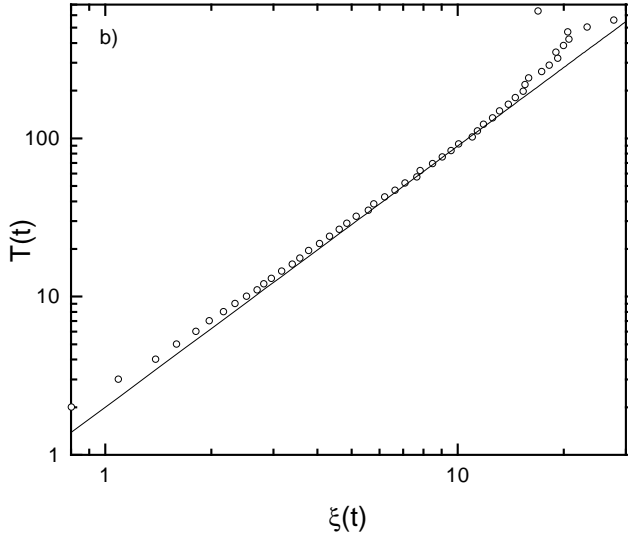}
\caption{Log-Log plot of the relation $T \sim \xi^z$ 
in $d=2$ (a) and $d=3$ (b). The full line 
corresponds to the value reported in Table II.}
\label{fig2}
\end{figure}

We now turn to the calculus of critical exponents. First we 
consider the exponent $z$ as defined in (\ref{def_z}). This was
computed by plotting the average duration of the avalanches
as a function of their characteristic average lengths. 
A binning procedure analogue to the one used in 
\cite{pra45.8551}
was employed. Plots are shown in Fig. \ref{fig2}.
Our best fit estimates are $1.34 \pm 0.02$ and $1.65 \pm 0.02$
in $d=2,3$ respectively, compatible with the BTW values
which are $4/3$ and $5/3$.
Remarkably, these results are also in perfect agreement 
with the RG results of Ref. \cite{Corral97}
which are $1.36 \; (d=2)$ and $1.68 \; (d=3)$.
The RG analysis was performed
on a Langevin equation where the driving
and the relaxation time scales are {\it comparable} (and
hence not well separated). Furthermore the strong (infinite)
non-linearity appearing in the continuum analogue of Eq. (\ref{relaxation_slow}),
was regularized and the result was analyzed within a one-loop RG 
scheme.
In view of all these approximations, the aforementioned closeness 
in the
two results is rather surprising.  We shall come back to this 
issue later on.

Another interesting critical exponent is the avalanche exponent 
size $\tau_s$ 
defined by the relation
\begin{eqnarray} \label{p(S)}
p(S) &=& S^{-\tau_s} F(S/L^{\phi}).
\end{eqnarray}
Here $p(S)$ is the distribution density of the avalanche sizes $S$,
$\tau_s$ is the avalanche exponent and $F(x)$ is a finite size
function defining the exponent $\phi$ \cite{pra39.6524}.
The function $F(x)$ is assumed to go to a constant
for small arguments (i.e. large sizes $L$) and to
"regularize" the large avalanche behaviour.
In order to improve the numerical estimates it proves convenient to look
at the integrated distribution density defined as 
\begin{eqnarray} \label{P(S)}
P(S) &=& \int_{0}^S \; ds \;\; p(s).
\end{eqnarray}
We have estimated the values of $\tau_s$ in two different ways. 
By plotting
the local slope (Fig.~3) and upon a finite size
procedure (analogue to the one used in Refs.~\cite{Lubeck97} 
and \cite{Lubeck97b}, see
Fig.~4). Both procedures yield consistent
results.  In $d=2$ our best estimate is $1.288 \pm 0.019$
which is close to the one given 
in Ref \cite{Lubeck97} by L\"{u}beck,
who reports $1.282 \pm 0.010$. It appears however that
the two extrapolations are not identical, since in his
analysis the values are {\em increasing} as $L$ increases
rather than {\em decreasing} as one would expect from a
finite size scaling.

Remarkably, both values are in good agreement with the BTW
value, thus supporting the claim that
the Zhang model belongs to the BTW universality class
\cite{Lubeck97}.
Our $d=3$ result is $1.454 \pm 0.041$ and it supersedes 
the one reported by Janosi \cite{pra42.769} namely $1.55$ 
which was presumably based only on small sizes 
(and thus too high according with
our previous discussion). However this disagrees with
the BTW value $4/3$ (see e.g. \cite{Lubeck97b})
and hence with the claim that the BTW and Zhang
model belong to the same universality class.

The values of $\phi$ were
computed from the collapse of the curves obtained
plotting $S^{\tau_s}\;P(S)$ versus $S/L^{\phi}$, that is
the universal finite size function. We find the best collapse
for $1.80 \pm 0.05$ and $2.6 \pm 0.1$. 
The error bars are estimated graphically.
A consistent value can be estimated by plotting
the size of the maximum avalanche as a function of the
size $L$, which is expected to scale as:
\begin{eqnarray} \label{smax}
s_{\mbox{\scriptsize max}} &\sim & L^{\phi}.
\end{eqnarray}
A log-log linear fit yields $1.84 \pm 0.06$ ($d=2$) and
$2.54 \pm 0.09$ ($d=3$).
A summary of all these critical exponents is reported
in Tables I and II.

Let us now turn to the behavior of the
roughness as defined in Eqs. (\ref{eqs:scaling}). 
As mentioned earlier, the dynamical
exponent $z$ can be found from the scaling {\it ansatz}
(\ref{def_z}). However, as it is usually done in the
field of growth processes \cite{barabasi}, one might
think to derive it from the scaling of the roughness as well.
In Fig ~\ref{fig5} we plot the roughness as defined
in (\ref{roughness_fast}).
We find (\ref{scal_w_fast}) to hold true with
$\beta_f=0.282 \pm 0.013$ and $\beta_f=0.391 \pm 0.031$
in $d=2,3$ respectively. These values were obtained
upon using an analysis similar to the one exploited
to compute $\tau_s$. From the collapse plot one can then
infer the value of $z_f$ appearing in (\ref{scal_w_fast}).
We find $z_f=1.20 \pm 0.05$ $(d=2)$
and $z_f=1.4 \pm 0.1$ $(d=3)$,
which are both lower than the corresponding value
derived from (\ref{def_z}). 
Similarly to (\ref{smax}) we have that 
\begin{equation}
T \sim L^{z_f}
\label{taufdef}
\end{equation}
with $z_f=1.19 \pm 0.04$
($d=2$) and $z_f=1.34 \pm 0.04$ ($d=3$).
Commonly the equality $z=z_f$ is tacitly assumed to hold and
we are not aware of any other examples where this point
was sufficiently emphasized.
A simple argument can be given here to explain this 
discrepancy. In usual interface growth phenomena the dynamical 
exponent
is measured as the scaling of the saturation time with the system 
length,
and this saturation occurs when the correlation length reaches the
system length. In our case, both lengths do not scale linearly but 
as 
$\xi\sim\L^{\eta}$. Thus these exponents need not be identical  
unless
$\eta=1$.  By a direct measurement (looking on how the maximum
$\xi$ scales with $L$)
we have found that 
$\eta=0.922\pm 0.012$ and 
$\eta=0.897\pm 0.051$, for $d=2$ and $3$, respectively. 
According to these scaling arguments we find that the product $z\eta$ 
agrees, within error bars, with the values reported for $z_f$.
In certain surface growth models a similar phenomenon, called 
anomalous scaling, has been reported \cite{juanma}. 
There it has been observed 
that the roughening exponents are different when measuring the 
local or the global widths.

Another exponent is derived from the relation $\chi_f=\beta_f\;z_f$
which is telling that the roughness, after that the avalanche
has been completed (i.e. at time $T(t)$), decreases as $L^{-\chi_f}$.
The values $\chi_f=\beta_f \; z_f$, according to
our previous results, are
$0.33$ and $0.55$ in $d=2,3$ respectively.
We now go back to the comparison with the RG results.

As previously hinted, although the exponent
$z$ derived from (\ref{def_z}) is very close to the
one derived by RG methods on the continuum Langevin
analogue of the Zhang model \cite{f1.963}, the 
$\beta_f$ and $\chi_f$ exponents are not.
A summary of all these values is reported in Table III
for compactness.  We argued previously that this inconsistency is 
not
surprising in view of the different physical regimes 
probed by the two cases and of the heavy approximations
involved in the RG calculation.
The apparent equality in the dynamical exponent $z$
then  probably hinges on deeper and more interesting
reasons, and we are planning to consider this in a
future work.

Finally, we have also measured the roughness on the
{\it slow} time scale as defined by (\ref{roughness_slow}).
We find that after a transient scaling as $L^d$, the
roughness tends to a limit which is {\it independent }
on $L$ (see Fig.~\ref{fig6}), i.e. eq. (\ref{scal_w_slow}) holds
with $\beta_s=0$, $\chi_s=0$ and $z_s=d$.

\section{Conclusions}
\label{sec:conclusions}
In this paper we have studied the infinitely slowly
driven Zhang model in two and three dimensions.
In two dimensions this work can be seen as
a complement of an earlier large sizes study \cite{Lubeck97}.
On the other hand in three dimensions, our results are expected to
improve an earlier estimate\cite{pra42.769}. The aim of 
\cite{pra42.769}
was different from ours and this could account for the difference.
In both cases we computed some exponents (notably the $\phi$ and all
the roughness exponents) which were never previously considered.
Besides being an useful complement to the existing literature
on the model we also found few unexpected results:
i) the three dimensional avalanche size
exponent does not
coincide with the BTW value, as the 
two-dimensional value seems to suggest;
ii) the exponent $z$ computed from
the dynamical scaling {\it ansatz} does not coincide with
the one computed from the roughening exponent.
We have shown that this stems from the non-linear scaling
of the correlation length $\xi$ with the system
size $L$;
iii)
the coincidence between the value of $z$
of the Zhang model and the RG value derived on
its Langevin continuum counterpart, does not extend
to other exponents such as the $\beta$ and the $\chi$
exponent.

We believe that all the above issues deserve further attention
both from the analytical and numerical view point.
We are currently performing a numerical investigation on the 
continuum
Langevin equation. This further analysis is expected to
shed new lights on the approximations involved in
the RG treatment.

\acknowledgments
Work in Italy has been supported by the Italian MURST (Ministero
dell'Universit\`{a} e della Ricerca Scientifica though the INFM
(Istituto Nazionale di Fisica della Materia). 
Work in Spain has been supported by
CICyT of the Spanish Government, grant \#PB94-0897.
Research was also partially supported by the Human Capital and 
Mobility
Programme, Access to Large Installations, under Contract no.
CHGE-CT92-009, "Access to Supercomputing Facilities for European
Researchers", established between the European Community, the 
Centre de
Supercomputacio de Catalunya and the Centre Europeu de 
Parallelisme de
Barcelona (CESCA/CEPBA).
Finally, we wish to thank G. Caldarelli, J.M. L\'{o}pez, and C. Tebaldi
for useful discussions. 


\begin{figure}
\epsfxsize=3.8truein\epsffile{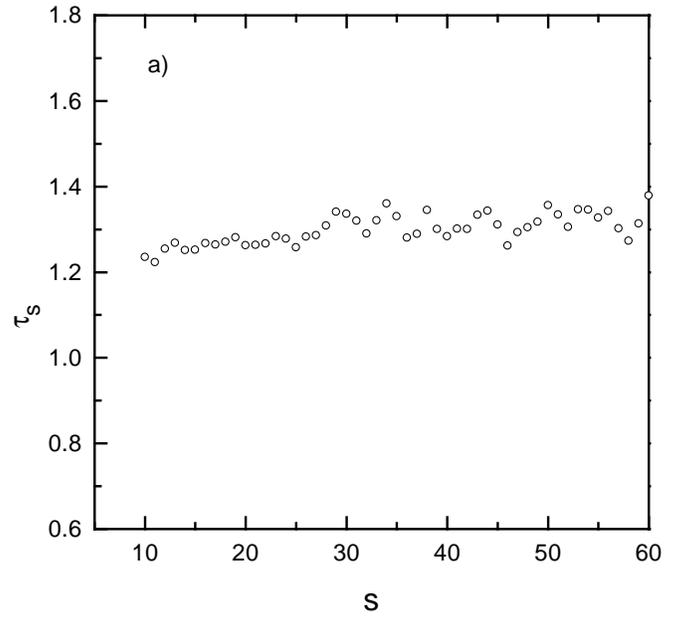}
\epsfxsize=3.8truein\epsffile{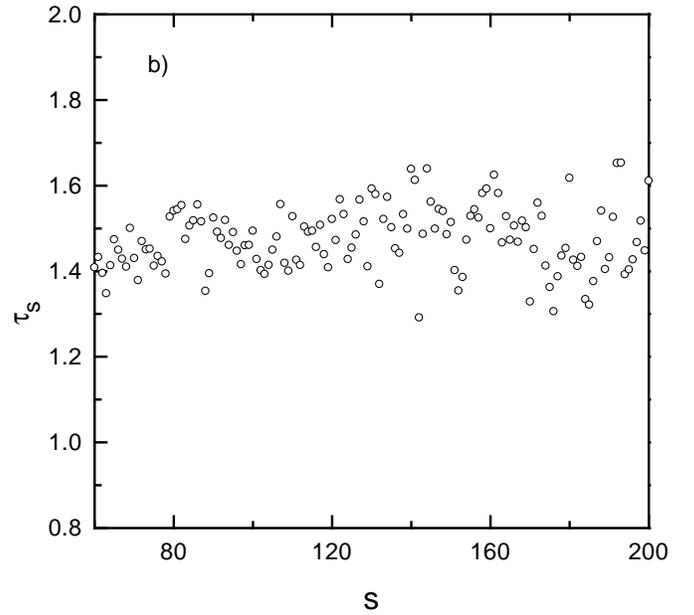}
\caption{Local slope plot for $\tau_s$ as a function of the
avalanche size $S$ in $d=2$ (a) and $d=3$ (b). In both cases
the intermediate most linear part of the largest size was
used for the computation.}
\label{fig3}
\end{figure}
\begin{figure}
\epsfxsize=3.8truein\epsffile{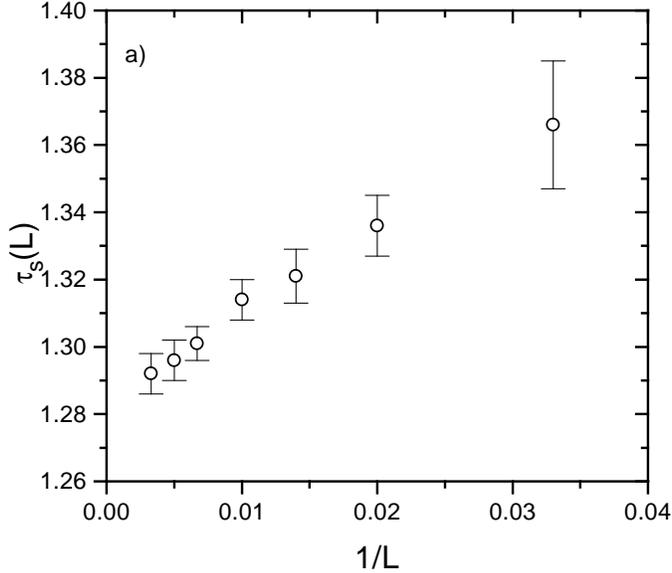}
\epsfxsize=3.8truein\epsffile{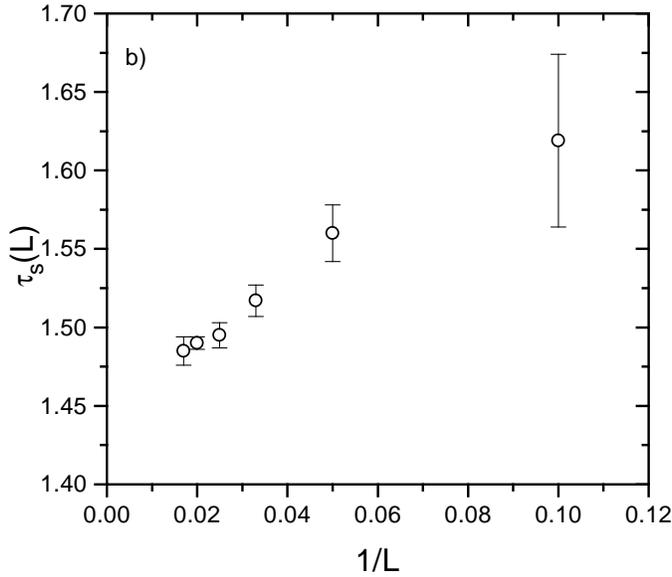}
\caption{Finite size plot for $\tau_s$ as a function of the
inverse lattice size $1/L$ in $d=2$ (a) and $d=3$ (b).}
\label{fig4}
\end{figure}
\begin{figure}
\epsfxsize=3.8truein
\epsffile{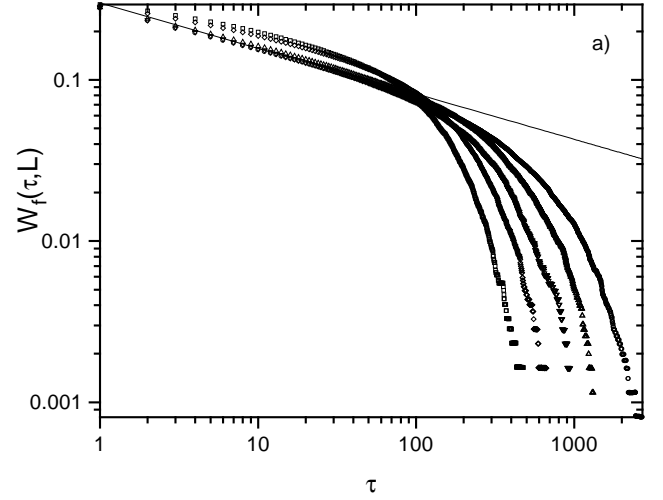}
\epsfxsize=3.8truein
\epsffile{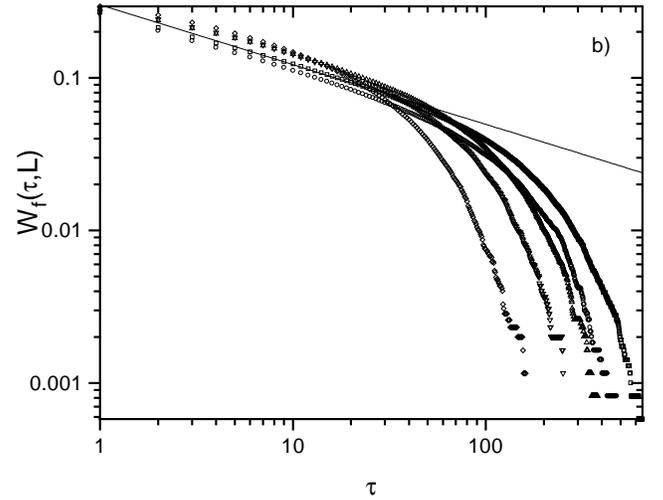}
\caption{Log-Log plot of $W_f(\tau,L)$ as function
of $\tau$ for various sizes $L$. In $d=2$ these were
$L$=70 ($\Box$), 100 ($\diamond$), 150 ($\nabla$),
200 ($\triangle$), 300 ($\circ$), and in $d=3$ they were
$L$=20 ($\diamond$), 30 ($\nabla$), 40 ($\triangle$),
50 ($\circ$), 60 ($\Box$). In both cases the solid line
corresponds to the value of $\beta_f$ reported in
Table III.}
\label{fig5}
\end{figure}
\epsfxsize=3.8truein\epsffile{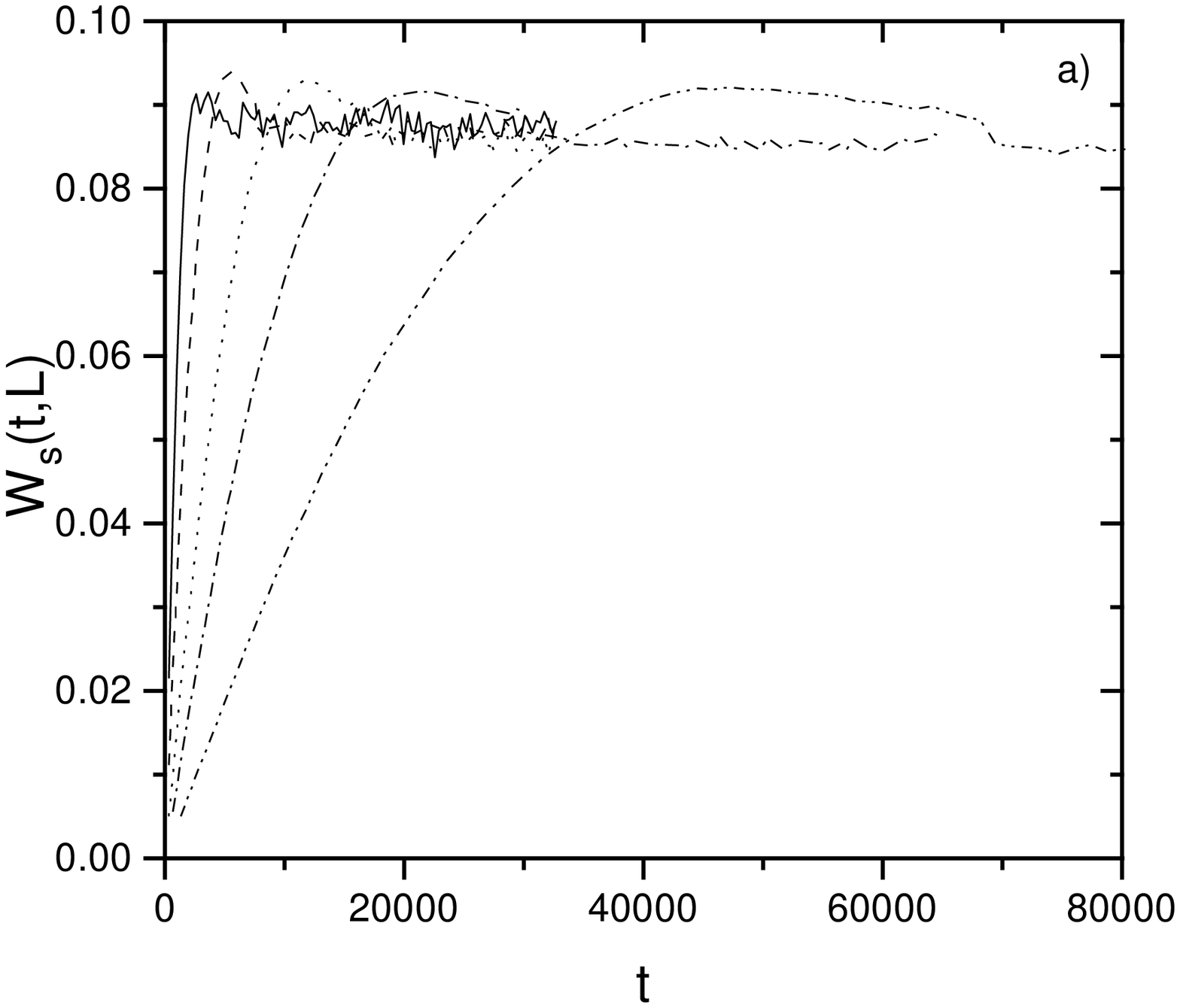}
\epsfxsize=3.8truein\epsffile{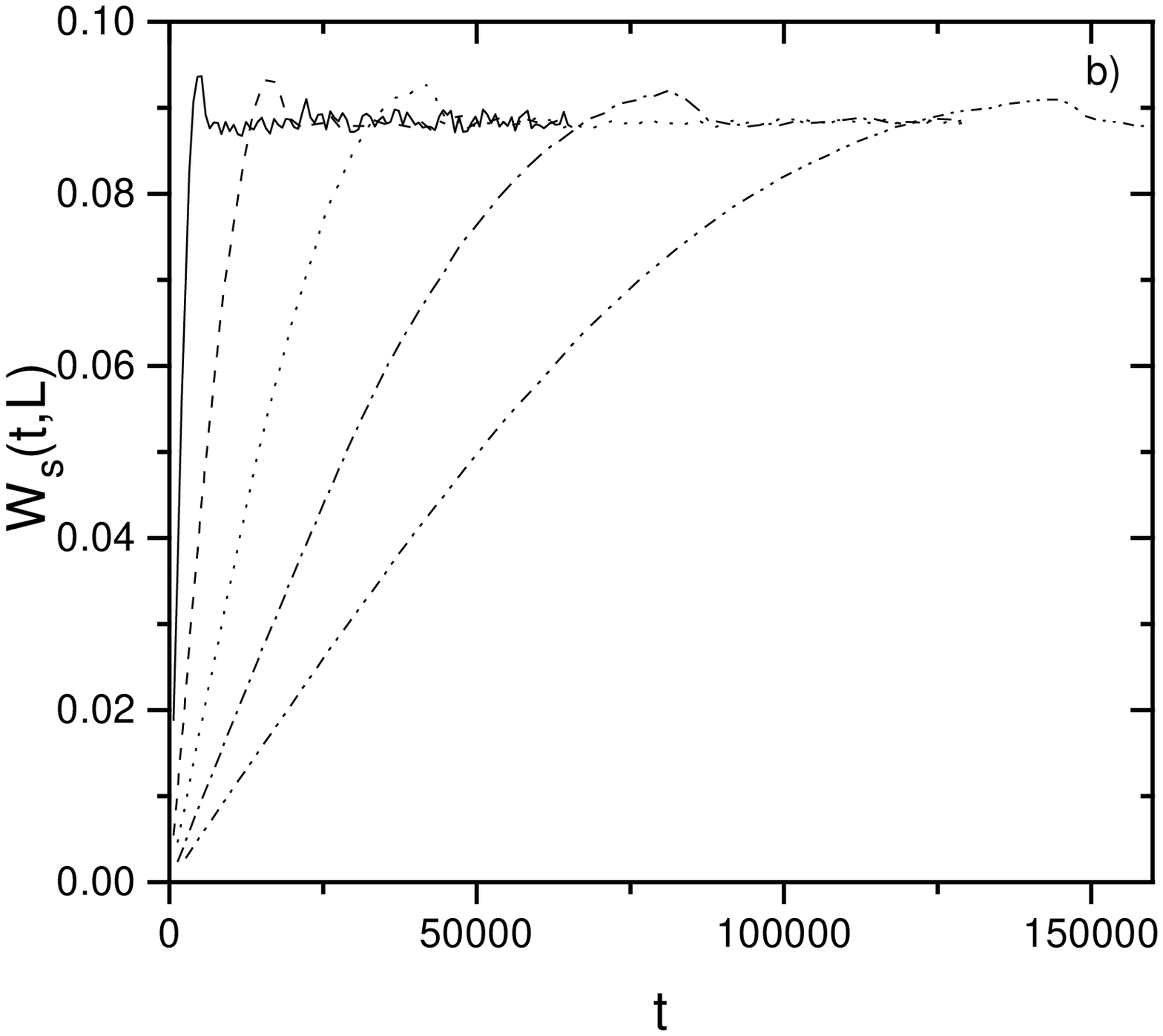}
\begin{figure}
\caption{Plot of $W_s(t,L)$ as function of $t$ for various
sizes $L$ both for $d=2$ (a) and $d=3$ (b). The values used
for the sizes are the same of the previous figure, the larger
$L$ the slower the growth.}
\label{fig6}
\end{figure}
\newpage
\begin{table}
\caption{Critical exponents $\tau_s$ and $\phi$ in $d=2,3$. 
The values indicated
by (a) and (b) refer to the BTW 
\protect{\cite{Lubeck97b}}
and the previous works 
\protect{\cite{Lubeck97,pra42.769}},
respectively. The
exponent $\phi$ given here is computed from 
(\protect{\ref{smax}}).} 
\begin{tabular}{lcccccc}
\multicolumn{1}{l}{$d$}&
\multicolumn{1}{c}{$\tau_s$}&
\multicolumn{1}{c}{$\tau_s(a)$}&
\multicolumn{1}{c}{$\tau_s(b)$}&
\multicolumn{1}{c}{$\phi$}&
\multicolumn{1}{c}{$\phi(a)$}&
\multicolumn{1}{c}{$\phi(b)$}\\
\hline
2 & $1.288 \pm 0.019$ & $1.293$ & $1.282 \pm 0.010$ &
$1.84 \pm 0.06$ & 2  & -  \\
3 & $1.454 \pm 0.041$ & 4/3 & 1.55 &
$2.54 \pm 0.09$ & 3  & -  \\
\end{tabular}
\label{table1}
\end{table}
\begin{table}
\caption{Dynamical critical exponent in $d=2,3$. The first column
corresponds to (5), whereas the second column is computed from (15).
Finally the last two columns
indicated by (a) and (b) are the
BTW 
\protect{\cite{Lubeck97b}} 
and the RG values 
\protect{\cite{f1.963}},
respectively.} 
\begin{tabular}{lcccc}
\multicolumn{1}{l}{$d$}&
\multicolumn{1}{c}{$z$}&
\multicolumn{1}{c}{$z_f$}&
\multicolumn{1}{c}{$z(a)$}&
\multicolumn{1}{c}{$z(b)$}\\
\hline
2 & $1.34 \pm 0.02$ & $1.19 \pm 0.04$ & 4/3 & 1.36 \\
3 & $1.65 \pm 0.02$ & $1.34 \pm 0.04$ & 5/3 & 1.68 \\
\end{tabular}
\label{table2}
\end{table}
\begin{table}
\caption{Roughness exponents $\beta$ and $\chi$ in $d=2,3$.
The values $\beta_f$ and $\chi_f=\beta_f \; z_f$ are computed 
here while
the others are the RG values \protect{\cite{f1.963}}.}
\begin{tabular}{lcccc}
\multicolumn{1}{l}{$d$}&
\multicolumn{1}{c}{$\beta_f$}&
\multicolumn{1}{c}{$\beta$}&
\multicolumn{1}{c}{$\chi_f$}&
\multicolumn{1}{c}{$\chi$}\\
\hline
2 & $0.282 \pm 0.013$ & -0.26    & $0.33 \pm 0.03$ & -0.36  \\
3 & $0.391 \pm 0.031$ & -0.1 & $0.55 \pm 0.08$ & -0.18 \\
\end{tabular}
\label{table3}
\end{table}

\begin{references}
\bibitem{note1}
In the original version 
\protect{\cite{prl63.470,pra45.8551}}
the driving systems was slightly different. Indeed
random site were successively selected and their
energy increased by a fixed amount. Relaxation
was taking place only when one of those sites
reached the critical threshold. We have tested
that, consistently with previous investigations,
this difference does not affect the critical exponents
and it speeds up the whole procedure.
\end{references}
\end{document}